\documentclass[a4paper,11pt]{article}
\pdfoutput=1 

\usepackage{jheppub}

\usepackage{amsmath,graphicx,amssymb,subfigure,bbm}
\usepackage[all]{xy}

\title{Backreacted flavor in non--commutative gauge theories}

\author[a,b]{Georgios Itsios,}
\author[c]{Veselin G. Filev,}
\author[d]{Dimitrios Zoakos,}
\affiliation[a]{Department of Engineering Sciences, University of Patras, 26110 Patras, Greece}
\affiliation[b]{Department of Mathematics, University of Surrey, Guildford GU2 7XH, UK}
\affiliation[c]{  School of Theoretical Physics, Dublin Institute for Advanced Studies, \\ 10 Burlington Road, Dublin 4, Ireland.}
\affiliation[d]{Centro de F\'isica do Porto \& 
Departamento de F\'isica e Astronomia, Faculdade de Ci{\^e}ncias \\ da Universidade do Porto, Rua do Campo Alegre 687, 4169--007 Porto, Portugal.}       
\emailAdd{gitsios@upatras.gr} \emailAdd{vfilev@stp.dias.ie}\emailAdd{dimitrios.zoakos@fc.up.pt}

\abstract{We construct the gravity dual of ${\cal N}=1$ Non-Commutative SYM theory coupled to $N_f$ smeared massless fundamental flavors.
Our solution is analytic and non-perturbative. Near the origin the background reduces to the one corresponding to an ordinary SYM theory. 
At large radial distances, the dilaton diverges signaling the presence of a UV Landau pole.  
Considering a probe D3--brane we calculate the effective YM coupling and show that it is independent on the parameter of non-commutativity. 
We calculate the corresponding beta function and show that it remains positive at all energy scales.}

\begin{document}
\begin{flushright}
DMUS-MP-13/10, DIAS-STP-13-05
\end{flushright}
\maketitle
\flushbottom


\section{Introduction}


Space-time non-commutativity (NC) is an idea introduced by Heisenberg and Pauli in the 1930, 
and since then investigated both from a physical and from a mathematical point of view. NC field theories have a built in minimal length scale, associated to the scale of non-commutativity. Despite the presence of an intrinsic UV cut off, non-commutative field theories still have quantum divergencies due to UV/IR mixing. 
However, the presence of minimal length scale agrees with the consensus that space-time should change its nature at microscopic scales comparable to the Plank length. Another important and relevant property of NC field theories is their nonlocality. As string theory is also non-local it is not a surprise that one of the most understood NC field theories are realized in its framework. In this paper we will study such a NC field theory in the context of the AdS/CFT correspondence.

Gauge/gravity correspondence \cite{Maldacena:1997re} relates the strongly coupled regime of a gauge theory with the weakly coupled regime of a string theory and vice versa. This duality puts a novel perspective in the study of strongly coupled gauge theories, beyond the ordinary perturbative analysis of a quantum field theory. In its original formulation the AdS/CFT correspondence relates ${\cal N}=4$ SYM in $1+3$ dimensions to superstring theory on the background of AdS$_5\times S^5$ space-time. 

Since the gauge theories on NC geometries may come out of certain limits of string theory 
\cite{Connes:1997cr, Seiberg:1999vs}, they have been the subject 
of intense study over the last years. Considering a system of D-branes with a constant  Neveu-Schwarz (NS) B field along the field theory directions, 
will couple open and closed strings. Performing a specific limit, it is possible for the closed strings to decouple and then the open string action corresponds to a NC gauge theory.  
In \cite{{Hashimoto:1999ut},{Maldacena:1999mh}} the correspondence was extended to NCSYM theories, by constructing the dual supergravity geometry 
related to D3--branes with an NS B-field.
In \cite{Witten:1985cc}, NC geometry has been the framework of analyzing open string field theory.


A major limitation of the original formulation of the AdS/CFT correspondence and its extension to NCSYM theory is that the field content of the dual gauge theory has only adjoint degrees of freedom. An important generalization of the correspondence was the addition of flavor degrees of freedom  \cite{Karch:2002sh} through the inclusion of probe D7--branes, that occupy the gauge theory directions and extend along the holographic. The probe limit of the D7--branes is valid when their number  is much less than the number of the color D3--branes. In the dual gauge theory this corresponds to the quenched approximation when fundamental loops are suppressed. When the number of flavors becomes comparable to the number of colors, we can no longer ignore the back reaction of the flavor branes and a fully backreacted supergravity background is needed.


Generally, it is very difficult to construct localized backreacted solutions, because of the breaking of rotational symmetry in the space transverse to the color branes. A way to circumvent this difficulty was developed in \cite{Bigazzi:2005md, Casero:2006pt} 
(see \cite{Nunez:2010sf} for a review and \cite{smearing} for some more recent references), where the broken global symmetries are restored, by smearing the flavor branes. The smearing has the effect of separating the flavor branes in the transverse space and making massive the string excitations connecting the different flavor branes. As a result the flavor symmetry of the dual gauge theory is $U(1)^{N_f}$ rather than $U(N_f)$, which would be the case for a localized solution. Also a smeared unquenched supergravity solution is generally less supersymetric, than a localized one. However, because of the enhanced global symmetry, the smeared solutions are much simpler and in some cases (as in our paper) analytic.


In this paper we construct the gravity dual of a non-commutative gauge theory coupled to $N_f$ {\it massless} fundamental flavors.
This corresponds to a Neveu-Schwarz (NS) B-field directed along the non-commutative directions and coupled to the fundamental degrees of 
freedom of the dual gauge theory. The presence of the NS B-field breaks the four dimensional Lorentz invariance, since it separates 
the {\it non-commutative} coordinates from the commuting ones. The background without flavors in \cite{Hashimoto:1999ut, Maldacena:1999mh}  is
coming from the decoupling limit of the type IIB supergravity solution of \cite{Russo:1996if, Breckenridge:1996tt, Costa:1996zd}, 
which in turn represents a stack of non-threshold D3 and D1-brane bound states. Adding a large number of massless flavor D7-branes, homogeneously
distributed over the internal space, will create a new type IIB supergravity solution with ${\cal N}=1$ supersymmetry. The construction we put forward with the current paper is 
complementary to the non-supersymmetric solutions presented in \cite{Filev:2011mt, Erdmenger:2011bw, Ammon:2012qs}, where the gravity dual of the (finite temperature) 
$SU(N_c)$  ${\cal N}=4$ SYM coupled to $N_f$  massless/massive fundamental flavors in the presence of an external magnetic field was derived. 


An overview of the paper is as follows: In section \ref{section:2} we present the ansatz, appropriate for the supergravity construction we previously discussed.
After substituting this ansatz to the known supersymmetric transformations for the dilatino and gravitino, we obtain a BPS system of ordinary 
differential equations together with the killing spinor. The background is $1/8$ supersymmetric. 


In section \ref{section:3} we solve the BPS system analytically and obtain expressions for every function of the background in terms of the 
holographic coordinate.  We analyze the solution both in the UV \& IR regimes. In the UV there is a Landau pole 
and a curvature singularity. 
In the IR there is also a curvature singularity (since the massless flavors cannot decouple even in the deep IR), 
but its character permits for field theory conclusions. This singularity is an artifact of the smearing procedure and disappears if one give a small mass to the flavor branes. 
To determine the radius/energy relation we calculate the energy of an open string stretching radially between the Landau pole and the IR. To calculate the effective Yang-Mills coupling we follow the approach of \cite{Seiberg:1999vs} and consider a probe D3--brane at fixed radial distance and extended along the field theory directions. Comparing the corresponding DBI action to an effective BI of the non-commutative SYM theory we show that the effective YM coupling is independent on the parameter of non-commutativity and coincides with the expression in the ordinary case. We calculated the corresponding beta function and showed that it is positive at all energy scales, which is consistent with the presence of a Landau pole and the vanishing of the gauge coupling in deep IR.


In appendix \ref{BPSanalysis} we put all the details of the computation of the BPS system that appears  in section \ref{section:2}, while 
in appendix \ref{EOMs} we list the second order differential equations, that the BPS system satisfies.


\section{Ansatz and the BPS equations} \label{section:2}

The ansatz we will adopt for the metric (in the Einstein frame) is \cite{Bigazzi:2009bk, Filev:2011mt, Erdmenger:2011bw, Ammon:2012qs, Benini:2006hh}
\begin{equation} \label{10dmetric}
ds_{10}^2 = h^{-\frac{1}{2}}\Big[-dt^2 + dx_1^2+b(dx_2^2+dx_3^2)\Big] + h^\frac{1}{2}
\Big[ b^2 S^8F^2 d\sigma^2 + S^2 ds_{CP^2}^2 + F^2 (d\tau + A_{CP^2})^2 \Big] \, ,
\end{equation}
where the $CP^2$ metric is given by
\begin{eqnarray}
ds_{CP^2}^2&=&\frac{1}{4} d\chi^2+ \frac{1}{4} \cos^2 \frac{\chi}{2} (d\theta^2 +
\sin^2 \theta d\varphi^2) + \frac{1}{4} \cos^2 \frac{\chi}{2} \sin^2 \frac{\chi}{2}(d\psi + \cos \theta d\varphi)^2
\quad \& \nonumber \\ [3pt]
A_{CP^2}&=& \frac12\cos^2 \frac{\chi}{2}(d\psi + \cos \theta d\varphi)\,\,.
\label{cp2metric}
\end{eqnarray}
The range of the angles is $0\leq (\chi, \theta) \leq \pi$,  $0\leq \varphi, \tau < 2\pi$, $0\leq \psi< 4 \pi$.
As usual the background is supplemented with a RR five-form
\begin{equation}  \label{F5}
F_5\,=\,K \,dt \wedge dx_1 \wedge dx_2 \wedge dx_3 \wedge d\sigma\,+\,{\rm Hodge\,\,dual} \, ,
\end{equation}
while the presence of a  RR one-form parametrizes the presence of a smeared D7-brane source in the system
\begin{equation} \label{F1}
F_1\,= \, Q_f \, (d\tau \, + \, A_{CP^2} ) \, .
\end{equation}
The background has  non-zero potentials $B_2$ and $C_2$ as well as non-zero RR 3-from:
\begin{equation} \label{NS+RR}
B_{2}  =  H dx^2\wedge dx^3 \, , \quad
C_{2} = J \,dt \wedge dx^1 \, , \quad \& \quad
F_{3} = d C_2\,+\, B_{2} \wedge F_1 \, .
\end{equation}
The potentials  $B_2$ and $C_2$ break the Lorentz symmetry along the field theory directions down to $SO(1,1)\times SO(2)$. This is reflected by the presence of the function $b$ in the metric \eqref{10dmetric}. 
In the dual field theory this breaking is due to the non-commutativity of the $(x_2,x_3)$ plane. 
All the functions of the ansatz, $h,b,S,F, \Phi, K, J \, \& \, H$, depend only on the radial coordinate. We choose the following frame for the metric  \eqref{10dmetric}
\begin{equation}
\arraycolsep=0.6cm
 \begin{array}{ll}
  e^0 \, = \,  h(\sigma)^{-{1 \over 4}} \, dt \, ,  & e^5 \, = \, \frac{1}{2} \, h(\sigma)^{1 \over 4} \, S(\sigma) \, d\chi \\ [3pt]
  e^1 \, = \,  h(\sigma)^{-{1 \over 4}} \, dx_1 \, ,  & e^6 \, = \, \frac{1}{2} \, h(\sigma)^{1 \over 4} \, S(\sigma) \, \cos\frac{\chi}{2} \, d\theta \\ [3pt]
  e^2 \, = \,  h(\sigma)^{-{1 \over 4}} \, b(\sigma)^{1 \over 2} \, dx_2 \, , & e^7 \, = \, \frac{1}{2} \, h(\sigma)^{1 \over 4} \, S(\sigma) \, \cos\frac{\chi}{2} \, \sin\theta \, d\phi \\ [3pt]
  e^3 \, = \,  h(\sigma)^{-{1 \over 4}} \, b(\sigma)^{1 \over 2} \, dx_3 \, ,  & e^8 \, = \,  \frac{1}{2} \, h(\sigma)^{1 \over 4} \, S(\sigma) \, \cos\frac{\chi}{2} \, \sin\frac{\chi}{2} 
 \left(d\psi \, + \,  \cos\theta \, d\phi\right)\\ [3pt]
  e^4 \, = \,  h(\sigma)^{1 \over 4} \, b(\sigma) \, S(\sigma)^4 \, F(\sigma) \, d\sigma \, , & e^9 \, = \,  h(\sigma)^{1 \over 4} \, F(\sigma) \, \left(d\tau \, + \,  A_{CP^2}\right) \, . 
 \end{array}
\end{equation} 
It is also useful to express the various differential forms in terms of the frame
\begin{eqnarray} \label{H3F1F3frame}
&&  H_3 \, = \,  \frac{H^{'}(\sigma) \, h(\sigma)^{1 \over 4}}{b(\sigma)^2 \, S(\sigma)^4 \, F(\sigma)} \, \, e^2 \wedge e^3 \wedge e^4 \,, \quad  
F_1 \, = \, \frac{Q_f}{h(\sigma)^{1 \over 4} \, F(\sigma)}\,\, e^9 \nonumber \\[3pt]
&&  F_3 \, = \, \frac{J^{'} \, h(\sigma)^{1 \over 4}}{b(\sigma) \, S(\sigma)^4 \, F(\sigma)} \, \, e^0 \wedge e^1 \wedge e^4  \, + \, 
\frac{Q_f \, H(\sigma) \, h(\sigma)^{1 \over 4}}{b(\sigma) \, F(\sigma)} \,\,e^2 \wedge e^3 \wedge e^9 \, .
\end{eqnarray}
The equation of motion for the five-form  implies immediately that
\begin{equation} \label{F5eom}
dF_5 \, = \, 0 \quad \Rightarrow \quad {K(\sigma) \,  h(\sigma)^2 \, b(\sigma)^{-2}} \,=\,{\rm constant}\,=\,Q_c \,,
\end{equation}
where the constant is obtained by imposing the Dirac quantization condition on the D3-brane charge. 
The constants $Q_{c}$ \&  $Q_{f}$ are proportional to the number of colors and flavors
\begin{equation}
N_c = \frac{Q_c\, Vol(S_5)}{(2\pi)^4g_s \,\alpha'^2} \quad \& \quad
N_f = \frac{4\,Q_f\,Vol(S_5)}{2 \, \pi^2 \, g_s} \, .
\end{equation}
The five-form in frame components is 
\begin{equation} \label{F5frame}
F_5 \, = \, {Qc \over F(\sigma) \, h(\sigma)^{{5 \over 4}} \, S(\sigma)^4} \, \Bigg[ e^0 \wedge e^1 \wedge e^2 \wedge e^3 \wedge e^4 \, 
- \,  e^5 \wedge e^6 \wedge e^7 \wedge e^8 \wedge e^9 \Bigg] \, .
\end{equation}
After an analytic derivation, which is presented in full detail in appendix 
\ref{BPSanalysis}, we come to the conclusion that the Killing spinor
in  the frame basis \eqref{10dmetric} can be written as 
\begin{equation} \label{spinor}
\epsilon \, = \,  h^{-\frac{1}{8}}\, e^{\frac{1}{2}\, \alpha \, \Gamma_{2\, 3 } \, \sigma_3}
\,  e^{-\frac{1}{2} \, i\, \sigma_2 \, \psi}
\,e^{-\frac{3}{2}\, i \, \sigma_2 \, \tau}\,\,\eta
\end{equation}
where $\eta$ is a constant spinor that satisfies the following set of projections
\begin{equation} \label{projections}
 i \, \Gamma_{0123}\,\sigma_2 \, \eta \, = \,  i \, \Gamma_{49}\,\sigma_2 \, \eta \, = \,  \eta \quad \& \quad 
i \, \Gamma_{58}\,\sigma_2 \, \eta \, = \, i \, \Gamma_{67}\,\sigma_2 \, \eta \, = \, - \, \eta \, ,
\end{equation}
while the angle $\alpha$ is defined through the following relation
\begin{equation} \label{tana}
\tan \alpha(\sigma) \, = \, \frac{H(\sigma) \, h(\sigma)^{1 \over 2}}{b(\sigma) \, e^{\frac{1}{2} \, \Phi(\sigma)}} \, .
\end{equation}
The analysis of the different components of the supersymmetric variations for the dilatino and gravitino, leads to the following BPS system of first-order differential equations
\begin{eqnarray}
\partial_{\sigma} \log b(\sigma) & = & J'(\sigma) \,  e^{\frac{1}{2} \, \Phi(\sigma)} \,  h(\sigma)^{1 \over 2} \sin \alpha \, - \, 
Q_f \, H(\sigma)\,  e^{\frac{1}{2} \, \Phi(\sigma)} \, S(\sigma)^4 \,  h(\sigma)^{1 \over 2} \sin \alpha \, ,  
\label{BPS-b}\\ [5pt]
\partial_{\sigma} \log h(\sigma) & = & {3 \over 2} \,  J'(\sigma) \,  e^{\frac{1}{2} \, \Phi(\sigma)} \,  h(\sigma)^{1 \over 2} \sin \alpha \, - \, {1 \over 2} \,  
Q_f \, H(\sigma)\,  e^{\frac{1}{2} \, \Phi(\sigma)} \, S(\sigma)^4 \,  h(\sigma)^{1 \over 2} \sin \alpha 
\nonumber \\ [5pt]
&-& Q_c \, b(\sigma) \, h(\sigma)^{-1} \, \cos \alpha \, ,
\label{BPS-h}\\ [5pt]
\partial_{\sigma} \log H(\sigma) & = &  Q_c \, b(\sigma) \, h(\sigma)^{-1} \, \cos \alpha  \, + \, 
Q_f \, H(\sigma)\,  e^{\frac{1}{2} \, \Phi(\sigma)} \, S(\sigma)^4 \,  h(\sigma)^{1 \over 2} \cot \alpha \cos \alpha \, ,
\label{BPS-H}\\ [5pt]
\partial_{\sigma}  \Phi(\sigma) & = & {1 \over 2} \Bigg[ \partial_{\sigma} \log b(\sigma) \, + \, 
2 \, Q_f \,  e^{\Phi(\sigma)} \,b(\sigma) \, S(\sigma)^4 \,  \cos^{-1} \alpha  \Bigg] \, ,
\label{BPS-Phi}\\ [5pt]
\partial_{\sigma}  J(\sigma) & = & - \, Q_c \, H(\sigma) \, e^{- \, \Phi(\sigma)} \,  h(\sigma)^{-1} \, , 
\label{BPS-J}\\ [5pt]
4 \, \partial_{\sigma} \log S(\sigma) & = &  4 \, b(\sigma) \,  S(\sigma)^2 \,  F(\sigma)^2 \, - \, 
 J'(\sigma) \,  e^{\frac{1}{2} \, \Phi(\sigma)} \,  h(\sigma)^{1 \over 2} \sin \alpha
\nonumber \\ [5pt]
&+& Q_f \, H(\sigma)\,  e^{\frac{1}{2} \, \Phi(\sigma)} \, S(\sigma)^4 \,  h(\sigma)^{1 \over 2} \sin \alpha \, , 
\label{BPS-S}\\ [5pt]
4 \, \partial_{\sigma} \log F(\sigma) & = & 12 \, b(\sigma) \,  S(\sigma)^4  \,- \, 
8 \, b(\sigma) \,  S(\sigma)^2 \,  F(\sigma)^2 \, - \, 
2 \, Q_f \,  e^{\Phi(\sigma)} \,b(\sigma) \, S(\sigma)^4 \,  \cos \alpha
\nonumber \\ [5pt]
&-& Q_f \, H(\sigma)\,  e^{\frac{1}{2} \, \Phi(\sigma)} \, S(\sigma)^4 \,  h(\sigma)^{1 \over 2} \sin \alpha \, - \, 
J'(\sigma) \,  e^{\frac{1}{2} \, \Phi(\sigma)} \,  h(\sigma)^{1 \over 2} \sin \alpha \, .
\label{BPS-F}
\end{eqnarray} 
Taking the limit $b \rightarrow 1$, $H \rightarrow 0$ \& $J \rightarrow 0$  to the BPS system \eqref{BPS-b}-\eqref{BPS-F}, we obtain the equations 
for the backreacted AdS$_5\times S^5$ background  
%
without the presence of non-commutativity (see \cite{Benini:2006hh} \& \cite{Bigazzi:2009bk}). In the limit 
$N_f \rightarrow 0$, we obtain the equations for the non-commutative deformation of $AdS_5 \times S^5$ (see \cite{Hashimoto:1999ut}, \cite{Maldacena:1999mh} 
\& \cite{Arean:2005ar})\footnote{The supersymmetry of this background, in a different parametrization for the metric, appears in \cite{Camino:2001ti}}.

Our next task is to count the number of supersymmetries preserved by the spinor defined in \eqref{spinor}. 
This can be calculated by dividing  the number 32 by two for each independent algebraic projection imposed on the constant spinor $\eta$.
Looking at \eqref{projections} we have three independent projections, so our deformed background (non-commutativity \& flavors) 
preserves four supersymmetries generated by the Killing spinor \eqref{spinor}.   

Final test for the BPS system \eqref{BPS-b}-\eqref{BPS-F} is the full set of the ten-dimensional equations of motion. 
This can be performed even before obtaining the actual solution for the BPS system. Counting on the analysis of \cite{Filev:2011mt}, 
where the full system of equations of motion is obtained, we checked that the first-order BPS system together with the Bianchi identities 
implies the second order equations, namely Einstein, Maxwell \& dilaton. In the appendix \ref{EOMs} we list the full set of equations of 
motion for every function of the background, while the analytic derivation can be  found in \cite{Filev:2011mt}.


\section{The solution of the BPS system} \label{section:3}

We begin by defining the field $U(\sigma)$ as follows
\begin{equation}
U(\sigma) \equiv \tan^2\alpha(\sigma)=\frac{H(\sigma)^2\,h(\sigma)}{e^{\Phi(\sigma)}\,b(\sigma)^2}\ .\label{def-U}
\end{equation}
The equation of motion for $U(\sigma)$ can be easily obtained by combing the equations of motion for $b(\sigma)$, $h(\sigma)$, $H(\sigma)$ and $\Phi(\sigma)$. 
Furthermore using \eqref{def-U} $h(\sigma)$ can be eliminated in favor of $U$. Finally we use \eqref{BPS-J} and \eqref{tana} 
to eliminate $J'(\sigma)$ and $\alpha(\sigma)$ from the equations of motion of the other field and in this way we obtain
\begin{eqnarray}
\partial_{\sigma}\log U(\sigma)&=&Q_c\,e^{-\Phi}\, b^{-1}\,H^2\,U^{-1}\,(1+U)^{1/2}\,+ \,Q_f\,e^{\Phi}\,b\,S^4\,(1+U)^{1/2}\ ,\label{A:EOM-U}\\ [5pt]
\partial_{\sigma}\log b(\sigma)&=&- \,Q_c\,e^{-\Phi}\,b^{-1}\,H^2\,(1+U)^{-1/2}\,- \, Q_f\,e^{\Phi}\,b\,S^4\,U\,(1+U)^{-1/2}\ , \label{A:EOM-b}\\ [5pt]
\partial_{\sigma}\log H(\sigma)&=&Q_c\,e^{-\Phi}\,b^{-1}\,H^2\,U^{-1}\,(1+U)^{-1/2}\,+ \,Q_f\,e^{\Phi}\,b\,S^4\,(1+U)^{-1/2}\ , \label{A:EOM-H}\\ [5pt]
2\,\partial_{\sigma}\Phi(\sigma)&=&- \, Q_c\,e^{-\Phi}\,b^{-1}\,H^2\,(1+U)^{-1/2} \, + \, Q_f\,e^{\Phi}\,b\,S^4\,(2+U)\,(1+U)^{-1/2}\ , \label{A:EOM-Phi}\\ [5pt]
4\,\partial_{\sigma}\log S(\sigma)&=&Q_c\,e^{-\Phi}\,b^{-1}\,H^2\,(1+U)^{-1/2}+Q_f\,e^{\Phi}\,b\,S^4\,U\,(1+U)^{-1/2}+4\,b\,F^2\,S^2\, ,~~ \label{A:EOM-S} \\ [5pt]
4\,\partial_{\sigma}\log F(\sigma)&=&Q_c\,e^{-\Phi}\,b^{-1}\,H^2\,(1+U)^{-1/2}\, - \, Q_f\,e^{\Phi}\,b\,S^4\,(2+U)\,(1+U)^{-1/2}\nonumber\\ [5pt]
&+& 4\,b\,\left(3\,S^4 \, - \, 2\,F^2\,S^2\right)\, . \label{A:EOM-F}
\end{eqnarray}
Next we compare \eqref{A:EOM-U} and \eqref{A:EOM-b} as well as \eqref{A:EOM-U} and \eqref{A:EOM-H}. It is easy to see that
\begin{equation}
\frac{b'(\sigma)}{b(\sigma)} \,= \, - \, \frac{U'(\sigma)}{1+U(\sigma)}  \, , \qquad 
\frac{H'(\sigma)}{H(\sigma)} \, = \, \frac{U'(\sigma)}{U(\sigma)(1+U(\sigma))}\ . \label{A:bH in U}
\end{equation}
We can solve both equations in \eqref{A:bH in U}  for $b$ and $H$ in terms of $U(\sigma)$
\begin{equation} \label{Solution-b-H}
b(\sigma)=\frac{c_b}{1+U(\sigma)} \, , \qquad H(\sigma)=\frac{c_H\,U(\sigma)}{1+U(\sigma)}\ , 
\end{equation}
where $c_b$ and $c_H$ are constants of integration to be determined later. The next step is to introduce a new radial variable $\rho$ and new fields defined as
\begin{eqnarray} \label{definitions}
&& \frac{d\rho}{d\sigma}\, =b\,S^4 \, = \, \frac{c_b\,S(\rho)^4}{1+U(\rho)} \, , \quad 
W(\rho) \, = \, \frac{S(\rho)^4}{1+U(\rho)} \, , 
\nonumber \\ [3.5pt]
&&
V(\rho) \, = \, \frac{F(\rho)^2}{S(\rho)^2} \, , \quad \& \quad 
Z(\rho)=e^{\Phi(\rho)}\,[1+U(\rho)]^{1/2} \, .
\end{eqnarray}
Using \eqref{A:bH in U} and after a bit of algebra we can obtain the following equations of motion for $U(\sigma)$, $W(\sigma)$, $V(\sigma)$ and $Z(\sigma)$
\begin{eqnarray}
\partial_{\rho}\log U(\rho)&=&Q_c\,c_H^2\,U\,(c_b^2\,W\,Z)^{-1}+Q_f \, Z \ , \label{EOMr-U}\\  [3.5pt]
\partial_{\rho}\log W(\rho)&=&4\, V\ , \label{EOMr-W}\\  [3.5pt]
\partial_{\rho}\log V(\rho)&=&6\,(1-V)-Q_f\,Z\ , \label{EOMr-V} \\  [3.5pt]
\partial_{\rho}\log Z(\rho)&=&Q_f\,Z \ . \label{EOMr-Z}
\end{eqnarray}
One can see that the equation of motion for $Z(\sigma)$  decouples and can be easily integrated
\begin{equation}
Z(\rho) \, = \, \frac{Z_*}{1 \, + \, \epsilon_* \, \left(\rho_* \, - \, \rho \right)} \quad {\rm with} \quad \epsilon_* \, = \, Q_f \, Z_*\ , \label{Solution-Z}
\end{equation}
where $\rho_*$ is a radial scale and the constant of integration has been fixed so that $Z(\rho_*) =Z_*$. 
We have also introduced the parameter $\epsilon_*$ in complete analogy with the  parameter $Q_f\,e^{\Phi_*}$ from the ordinary case \cite{Bigazzi:2009bk}. 
Note that $Z$ is related to the dilaton and has a pole at
\begin{equation}
\rho_{LP} \, = \, \frac{1}{\epsilon_*} \, + \, \rho_*\ . \label{LP}
\end{equation}
If the dilaton diverges at $\rho_{LP}$ (this is when  $\sqrt{1+U(\rho)}$ diverges slower than $Z(\rho)$ as $\rho\rightarrow \rho_{LP}$), we can think of $\rho_{LP}$ as the energy scale corresponding to the Landau pole of the theory. Such a behavior is expected due to the positive contribution of the flavor degrees of freedom to the beta function of the theory. 

\noindent In order to disentangle the rest of the equations, we combine (\ref{EOMr-W}), (\ref{EOMr-V}) and (\ref{EOMr-Z})
\begin{equation}
\partial_{\rho}\log(Z\,V\,W^{3/2}) \, = \, 6\ , \label{EOM-ZVW}
\end{equation}
and solve (\ref{EOM-ZVW}) for $V$ in terms of $Z$ and $W$
\begin{equation}
V(\rho) \, = \, c_V\, W(\rho)^{-3/2}\,Z(\rho)^{-1}\, e^{6\rho}\ . \label{EOM-VtoWZ}
\end{equation}
Substituting \eqref{EOM-VtoWZ} into the equation of motion for $W$ \eqref{EOMr-W} and using \eqref{Solution-Z}, 
we obtain a first order differential equation for $W$ which we can solve
\begin{equation}
W(\rho) \, = \, \alpha'^2\, e^{4\rho}\,\Bigg[1 \, + \, \epsilon_* \, \left(\frac{1}{6} \, + \, \rho_* \, - \, \rho  \right) \Bigg]^{2/3} \ ,\label{Solution-W}
\end{equation}
where we have fixed the constant of integration in complete analogy to the ordinary case \cite{Benini:2006hh}. The corresponding solution for $V(\rho)$ is
\begin{equation}
V(\rho) \, = \, \frac{1 \, + \, \epsilon_* \, \left(\rho_* \, - \, \rho \right)}{1 \, + \, \epsilon_* \, \left(\frac{1}{6} \, + \, \rho_* \, - \, \rho \right)}\, . \label{Solution-V}
\end{equation}
Our next step is to substitute the solutions for $Z,V$ and $W$ from equations (\ref{Solution-Z}), (\ref{Solution-W}) and (\ref{Solution-V}) into the equation of motion for $U$ (\ref{EOMr-U}). The resulting differential equation for $U$ is
\begin{equation}
U'(\rho) \, = \, \frac{Q_c\,c_H^2}{\alpha'^2\,c_b^2\,Z_*}\,\frac{1\, + \, \epsilon_* \left(\rho_*-\rho\right)}{1+\epsilon_* 
\left(\frac{1}{6} \, + \, \rho_*-\rho\right)} \, U(\rho)^2 \, + \, \frac{\epsilon_*}{1 \, + \, \epsilon_*(\rho_* \, - \, \rho)} \, U(\rho)\ ,
\end{equation}
which can be easily solved to give
\begin{equation} \label{Solution-U}
U(\rho) \, = \, \alpha'^2\,\frac{2^{2/3}\,c_b^2\,e^{\frac{2}{3}+4\rho_{LP}}\,Z_*}{c_H^2\,Q_c\,\epsilon_*^{1/3}}\,
\frac{1}{\rho_{LP} \,- \, \rho}\,
\frac{1}{c_U \, + \, e^{\frac{2}{3}\,i\,\pi}\,\Gamma\left(\frac{1}{3},-\frac{2}{3} \, - \, 4 \rho_{LP} \, + \, 4 \rho, \, 0 \, \right)} \,  ,
\end{equation}
where $\rho_{LP}$ is the scale of the Landau pole defined in (\ref{LP}), $c_U$ is a real constant of integration and 
$\Gamma (a,z_1,z_2)  \equiv  \Gamma(a,z_1)  -  \Gamma(a,z_2)$ 
is the generalized incomplete gamma function. The last piece in the puzzle is to combine \eqref{Solution-W}, \eqref{Solution-V} \& \eqref{Solution-U} together
with \eqref{definitions} to obtain expressions for the $h, \Phi, F$ \& $S$ as functions of $\rho$
\begin{eqnarray} \label{Solution-h-phi-S-F}
&& h(\rho) \,  = \,  \frac{Z(\rho)}{c_H^2 \, U(\rho) \sqrt{1 \, + \,  U(\rho)}} \, , \quad e^{\Phi(\rho)} \, = \, \frac{Z_*}{1 \, + \, \epsilon_* \, \left(\rho_* \, - \, \rho \right)}\, 
 \frac{1}{\sqrt{1 \, + \,  U(\rho)}} \, , 
\nonumber \\ [5pt]
&&  S(\rho)  \, =  \,  \sqrt{\alpha'} \,  \left(1 \, + \,  U(\rho) \right)^{1/4} \, e^{\rho}\,\Bigg[1 \, + \, \epsilon_* \, \left(\frac{1}{6} \, + \, \rho_* \, - \, \rho  \right) \Bigg]^{1/6} \, , 
\\ [5pt] 
&&  F(\rho)  \, =  \,  \sqrt{\alpha'} \,  \left(1 \, + \,  U(\rho) \right)^{1/4}  \, e^{\rho}\, 
\Big[1 \, + \, \epsilon_* \, \left(\rho_* \, - \, \rho \right) \Big]^{1/2}\, 
\Bigg[1 \, + \, \epsilon_* \, \left(\frac{1}{6} \, + \, \rho_* \, - \, \rho  \right) \Bigg]^{-1/3} \,.
\nonumber
\end{eqnarray}

An alternative way to obtain the same solution is through a TsT transformation \cite{Lunin:2005jy} on the  backreacted AdS$_5\times S^5$ background  
without the presence of non-commutativity \cite{Benini:2006hh, Bigazzi:2009bk}.
This transformation consists of a T-duality along the isometry direction $x_3$, 
a coordinate shift along $x_2 \rightarrow x_2 + \gamma x_3$, where the parameter $\gamma$ is related to the non-commutativity,  
and finally another T-duality along $x_3$. When the internal six dimensional (squashed after the smearing procedure) metric does not change under the {\it rotation} 
that the TsT duality performs, then smearing and TsT commute. This is a claim that previously appeared in more general framework in \cite{Gaillard:2010qg}.

The way to determine the integration constant $c_U$ is through the study of the poles of the function $U(\rho)$ 
(and consequently of the dilaton, through \eqref{definitions}). 
Demanding the denominator of the last fraction of \eqref{Solution-U} to vanish at $\rho = \rho_{LP}$, the function  $U(\rho)$  has a second order pole 
at this point. As we can see from \eqref{Solution-h-phi-S-F}, in this way the dilaton does not have a pole at $\rho = \rho_{LP}$. However the curvature invariants, 
in the string frame, diverge when $\rho \rightarrow \rho_{LP}$ and this is a sign that the supergravity approximation is not to be trusted in the UV. 
For example the Ricci scalar is $R \sim (\rho_{LP} - \rho)^{-1}$ . Demanding the denominator of \eqref{Solution-U} to vanish at $\rho = \rho_{1}$ with 
$\rho_1 \ne \rho_{LP}$, the function $U(\rho)$ has two single poles at $\rho = \rho_{LP}$ and $\rho = \rho_{1}$. 
Now there are two possibilities, either  $\rho_1<\rho_{LP}$ or  $\rho_1>\rho_{LP}$.

If $\rho_1<\rho_{LP}$, the background terminates at $\rho=\rho_1$ and there is a curvature singularity there.
This case is unphysical since the singularity does not correspond to a Landau pole (the dilaton remains finite at $\rho=\rho_1$).

If $\rho_1>\rho_{LP}$, the geometry terminates at $\rho=\rho_{LP}$ and only the first pole of $U(\rho)$ is realized. 
One can check that the dilaton has a singularity at $\rho = \rho_{LP}$ ({\it milder} with respect to the ordinary case $\exp[\Phi] \sim (\rho_{LP} - \rho)^{-1/2}$), while the Ricci scalar diverges as $R \sim (\rho_{LP} - \rho)^{-5/2}$. The singularity at $\rho=\rho_{LP}$ is interpreted as a Landau pole (the dilaton diverges), 
thus the physically meaningful case is for $\rho_{LP} < \rho_{1}$.

The analysis above indicates that in order to have physically relevant solutions, the constant $c_U$ is restricted to 
\begin{equation}\label{eqCu}
c_U \, > \, c_U^{cr} \, \equiv \, -e^{\frac{2}{3}\,i\,\pi}\, \Gamma \left[ {1 \over 3}, - {2 \over 3}, 0\right]\approx -3.1555 \, ,
\end{equation}
where $c_U^{cr}$ correspond to the case when $\rho_1=\rho_{LP}$. We fix the constant $c_U$ to be zero, which is consistent with the range given in (\ref{eqCu}).

In the IR limit of the geometry $(\rho\rightarrow - \infty)$ the solution is independent from the effect of non-commutativity, while the effect of flavors 
in the Einstein frame is coming through subleading corrections to the leading term. 
The behavior of the solution when a large number of massless flavors backreact on the geometry is known from the analysis of \cite{Benini:2006hh}, 
while from \cite{Hashimoto:1999ut} and \cite{Maldacena:1999mh}  it is also known that the presence of a background B-field changes the dynamics of the 
closed strings only far away from the horizon. The geometry near the horizon in the Einstein frame is described by the usual $AdS_5 \times S^5$ solution with a constant curvature. In fact one can obtain 
\begin{equation}
R^{E} \sim 0 \quad \& \quad  R^{E}_{\mu \nu} \, R^{E,\mu \nu} \sim \frac{640}{Qc} \, 
\end{equation}
for the $\rho\to-\infty$ limit of the curvature invariants in the Einstein frame. However, the dynamics is not that of the IR limit of the usual AdS$_5\times S^5$ background with constant dilaton, because $e^{\Phi(\rho)}\to 0$ in the $\rho\to-\infty$ limit. This discrepancy is expected since our backreacted geometry corresponds to the introduction of massless flavors to the dual gauge theory and there is no reason to expect the flavors to decouple in the IR limit. In fact in \cite{Benini:2006hh} it was shown that the background has a curvature singularity in the string frame. Indeed, calculations of the Ricci scalar and the square of the Ricci tensor reveal the singularity in the IR of the theory
\begin{equation}
R \sim (- \rho)^{-1/3} \quad \& \quad  R_{\mu \nu} \, R^{\mu \nu} \sim (- \rho)^{8/3} \rightarrow \infty \, .
\end{equation}
This phenomenon is solely due to the presence of the backreacted massless flavors in the geometry, since the non-commutativity does not affect the IR properties of the theory.
According to the criterion that was put forward in \cite{Maldacena:2000mw} (see also  \cite{Gubser:2000nd}) an IR singularity is physically acceptable,  
if the metric component $g_{tt}$ is bounded near the problematic point. Since this is the case for our solution this is a  {\it good singularity}
and that means that we can extract gauge theory physics from these supergravity backgrounds. 

In fact this singularity can be regulated by assigning a small bare mass to the backreacted flavor branes. To this end we have to construct the geometry dual to NCSYM with massive flavors. While we leave this study for another project, we can still deduce a qualitative description from the study of massive backreacted  flavor branes in the ordinary case preformed in  \cite{Bigazzi:2008zt}. Indeed, in the IR limit the NCSYM reduces to an ordinary SYM, therefore if the mass of the flavors is sufficiently small the corresponding backreacted geometry should approach the solution of  \cite{Bigazzi:2008zt}. In this paper the authors showed that the effect of the finite massive flavors is to produce a spherical cavity of radius proportional to the mass of the flavors, while inside the geometry is given by the unflavored solution (AdS$_5\times S^5$ with constant dilaton in our case). 
The dilaton freezes at the cavity and remains constant inside, thus we avoid the curvature singularity in the string frame. 
The physical interpretation is that the massive flavors decouple in the IR.


\subsection{Limits of the geometry}

So far we have fixed some of the integration constants of our solution, either by analogy to the ordinary case of \cite{Benini:2006hh} and \cite{Bigazzi:2009bk} or by extracting information from the UV asymptotic. Let us know explore how our solution reproduces the unflavored solution in the limit of vanishing flavors $N_f\rightarrow 0$. Focusing on the expressions for the NS two-form, the function $b$ and the dilaton we take the limit $N_f\rightarrow 0$ in equations \eqref{Solution-b-H} and \eqref{Solution-h-phi-S-F}
\begin{equation} \label{H-noflavorlimit}
H \, \sim \, \frac{c_H}{1 \, + \, e^{-4 \rho} \, \frac{c_H^2 \, R^4}{c_b^2\,Z_* \, \alpha'^2}}\,  ,
\quad b\, \sim \, \frac{c_b}{1\,+\,e^{4\rho}\,\frac{c_b^2\,Z_*\,\alpha'^2}{c_H^2\,R^4}}\, ,
\quad e^{\Phi_s}=\, \frac{g_s\, Z_*}{\sqrt{1\,+\,e^{4\rho}\,\frac{c_b^2\,Z_*\,\alpha'^2}{c_H^2\,R^4}}}\, ,
\end{equation}
where $e^\Phi_s=g_s\,e^{\Phi}$ is the dilaton in the string frame. The non-commutative expressions for the  NS two-form, the function $b$ and the dilaton coming from previous studies in the literature (see e.g. \cite{Arean:2005ar} and \cite{Erdmenger:2011bw}) are\footnote{Note that we use a plus sign convention for the $B$-field, unlike the 
one used in \cite{Arean:2005ar}.}
\begin{equation} \label{H-noncommutative}
H \, = \,  \Theta^2 \, \frac{r^4}{R^2}\, \frac{1}{1 \, + \,  \Theta^4 r^4}\ , ~~~b=\frac{1}{1\,+\,\Theta^4\,r^4}\ ,~~~e^{\Phi_s}=g_s\,\frac{1}{\sqrt{1+\Theta^4\,r^4}} \, ,
\end{equation}
where $\Theta$ is the non-commutativity parameter.
Identifying \eqref{H-noflavorlimit} and \eqref{H-noncommutative}, with the proper change of coordinates, we are able to relate the constant $c_H$ to $\Theta$
\begin{equation} \label{def-theta}
r \, \rightarrow \,  \alpha'^{1/2} \, e^{\rho}\ ,\quad \Theta \, \rightarrow \, \frac{1}{\sqrt{c_H} \, R} \, \ ,\quad c_b\rightarrow1\, \ ,\quad Z_*\rightarrow 1\ .
\end{equation}
Note that at finite number of flavors the constants $c_b$ and $Z_*$ may be different from one. In the next subsection we will fix $c_b=1$, however the constant $Z_*$ will remain as a free parameter representing the fact that in the flavoured theory the gauge coupling (the dilaton) runs even in the IR limit. 

Using the last relation in equation (\ref{def-theta}) we see that the commutative limit $\Theta\rightarrow 0$ corresponds to taking the limit $c_H \rightarrow\infty$. Indeed, in this limit our solution,\eqref{Solution-b-H} and  \eqref{Solution-h-phi-S-F} reduces to the supergravity background dual to  ${\cal N} =1$ SYM with backreacted massless flavors appearing in \cite{Benini:2006hh} and \cite{Bigazzi:2009bk}.


\subsection{Choice of radial coordinate}

Inspired from the limit of vanishing flavors considered above we make a choice of radial coordinate $r$ in which the metric (in the string frame) along the filed theory directions $t,\vec x$ has the same form as in the unflavored case. This will help us to obtain an expression for the parameter of non-commutativity at finite number of flavors. Comparing equations  (\ref{Solution-b-H}) and (\ref{H-noncommutative}) we make the natural definitions
\begin{equation}\label{def-theta-r}
U(\rho)\,=\,\Theta^4\,r(\rho)^4\ , ~~~c_b=1\ ,~~~ \Theta \, =\, \frac{1}{\sqrt{c_H} \, R}\ .
\end{equation}
Note that the last expression is the same as in (\ref{def-theta}). Equations (\ref{Solution-U}) and (\ref{def-theta-r}) imply the following definition of radial coordinate $r(\rho)$
\begin{equation}\label{def-r}
r(\rho) \, = \, \alpha'^{1/2}\,Z_*^{1/4}\,\frac{e^{\frac{1}{6}+\rho_{LP}}}{\,2^{1/3}\,\epsilon_*^{1/12}}\,
\left({\rho_{LP}-\rho}\right)^{-1/4}\,\left(
{ \, e^{\frac{2}{3}\,i\,\pi}\,\Gamma\left(\frac{1}{3},-\frac{2}{3} \, - \, 4 \rho_{LP} \, + \, 4 \rho, \, 0 \, \right)} \right)^{-1/4}\,  ,
\end{equation}
where we have fixed $c_U=0$ and $\rho_{LP}$ is defined in equation (\ref{LP}). One can check that in the limit of vanishing number of flavors $\epsilon_*\to 0$ one recovers the expression $r \, \to\,  \alpha'^{1/2} \, e^{\rho}$ from equation (\ref{def-theta}).  Let us check that the definitions (\ref{def-theta-r}), (\ref{def-r}) render the metric along the field theory direction in the same form as in the unflavored case. Transforming the metric into the string frame and using equations (\ref{Solution-h-phi-S-F}) and (\ref{def-theta-r}) we obtain
\begin{equation}\label{G-string}
-G^{s}_{00}\,=\,G^{s}_{11}=\,c_H\,U(\rho)^{1/2}\,=\,\frac{r^2}{R^2}\ ,~~~G^s_{22}\,=\,G^s_{33}\,=\,b(\rho)\,G^s_{11}\,=\,\frac{G^s_{11}}{1\,+\,\Theta^4\,r^4}\ .
\end{equation}
which are the same as in the unflavored case (we use the conventions of ref.~\cite{Arean:2005ar} ).  Furthermore, the dilaton is given by
\begin{equation}\label{dilaton-string}
e^{\Phi_s}=\frac{g_s\,Z_*}{1+\epsilon_*(\rho_*-\rho(r))}\,\frac{1}{\sqrt{1+\Theta^4\,r^4}}\,=\,e^{\Phi_0(r)}\,\frac{1}{\sqrt{1+\Theta^4\,r^4}}\ ,
\end{equation} 
where $e^{\Phi_0}$ is the expression for the dilaton from the ordinary case ($\Theta=0$). Just like in the unflavored case the effect of the non-commutativity is to ``dress" the dilaton by the factor $1/\sqrt{1+\Theta^4\,r^4}$. The difference is that in the unflavored case $e^{\Phi_0}$ is constant, while the flavored theory is not conformal and $e^{\Phi_0}$ runs.


\subsection{Radius/energy relation}

The fact that the metric along the field theory directions has the same form as in the unflavored case may tempt us to associate the radial variable $r$ (defined in (\ref{def-r})) with the energy scale of the theory. However, in the unflavored case the metric has an AdS throat  and rescaling of the $(x_0,x_1)$ plane can be compensated by rescaling of the radial coordinate, which leads to the radius/energy relation. This is not the case for our backreacted supergravity background. An alternative route could be studying the space-like geodesics near the boundary, however our geometry has a curvature singularity at $\rho_{LP}$, we cannot trust our solution beyond this point and a UV boundary cannot be defined. Nevertheless we could still define $r$ to be proportional to the energy scale, after all this choice is scheme dependent. However, such a choice would assign an infinite energy scale to the Landau pole ($r(\rho_{LP})=\infty$), while one can check that the proper distance between a point in the bulk of the geometry and the Landau pole is finite. In fact we can estimate the energy scale of the Landau pole by calculating the energy of an open string stretched radially between the Landau pole ($\rho=\rho_{LP}$) and the IR ($\rho=-\infty$). Using the Nambu-Goto action we obtain
\begin{equation}
\Lambda_{LP}=\frac{1}{2\,\pi\,\alpha'}\,\int\limits_{-\infty}^{\rho_{LP}}\,d\rho\,\sqrt{-G_{00}^{s}\,G_{\rho\rho}^{s}}\,=\,\frac{1}{2\,\pi\,\alpha'}\,\int\limits_{-\infty}^{\rho_{LP}}\,d\rho\,e^{\frac{1}{2}\Phi(\rho)}\,F(\rho)\ .
\end{equation}
Using equations (\ref{Solution-h-phi-S-F}) we obtain
\begin{equation}
\Lambda_{LP}\,=\,\frac{e^{1/6}\,Z_*^{1/2}\,\Gamma\left(\frac{2}{3},\frac{1}{6}\right)}{2\,\pi\,\epsilon_*^{1/3}\,\sqrt{\alpha'}\,}\,e^{\rho_{LP}}\,\sim\,e^{\rho_{LP}}\ .
\end{equation}
Observing that $\Lambda_{LP}\sim e^{\rho_{LP}}$ we assign to the radial variable $\rho$ an energy scale $\mu\sim e^{\rho}$. The same choice of energy scale was used in ref.~\cite{Benini:2006hh}. The precise radius/energy relation is
\begin{equation}\label{radius/energy}
\rho =\rho_{LP}+\log\frac{\mu}{\Lambda_{LP}}\ . 
\end{equation}
Using equation (\ref{radius/energy}) we can calculate the running of the Yang-Mills coupling and the corresponding beta function.


\subsection{Effective YM coupling and Beta function}

In this subsection we study the effective Yang-Mills coupling $g_{YM}^2$ of the dual non-commutative field theory. In order to relate the coupling constant to the supergravity parameters we use the approach of  \cite{Seiberg:1999vs} In section 2 of the same paper, the authors adopt a point-splitting regularization to show that the opens string sector of a string theory with closed string parameters $g$ and $B$ in terms of non-commutative YM theory with parameter of non-commutativity is given by\footnote{Note that in our conventions the $B$-field of 
\cite{Seiberg:1999vs} is transformed to $B\to B/(2\pi\alpha')$.}
\begin{equation}\label{thet}
\theta^{ij}=-(2\,\pi\,\alpha')\,\left(\frac{1}{g+B}\,B\,\frac{1}{g-B}   \right)^{ij} \ .
\end{equation}
To bosonic part of the action of the non-commutative YM can be written as \cite{Seiberg:1999vs}
\begin{equation}
\frac{(\alpha')^{\frac{3-p}{2}}}{(2\pi)^{p-2}\, G_s}\,\int\,\sqrt{\det G}\,G^{ii'}\,G^{jj'}\,\frac{1}{4}\,{\rm{Tr}}\, \hat F_{ij}\ast \hat F_{ij}\ ,
\end{equation}
where $\ast$ is the Moyal product with $\theta$ given in (\ref{thet}). $G_s$ is the effective string coupling,
\begin{equation}
\hat F_{ij}=\partial_i\hat A_j-\partial_j\hat A_i-i\hat A_i\ast\hat A_j+i\hat A_j\ast\hat A_i
\end{equation}
is the non-commutative field strength and $G$ is the open string metric corresponding to $g$ and $B$ defined by
\begin{equation}
G^{ij}=\left(\frac{1}{g+B}\,g\,\frac{1}{g-B} \right)^{ij}\ ,~~~G_{ij}=g_{ij}-(B\,g^{-1}\,B)_{ij}\ .
\end{equation}
For slowly varying fields the non-commutative action can be described by BI action
\begin{equation}\label{BI action}
{\cal L}(\hat F)=\frac{1}{G_s(\mu)\,(2\pi)^p\,(\alpha')^{\frac{p+1}{2}}}\,\sqrt{\det (G+2\pi\alpha'\,\hat F)}\ ,
\end{equation}
where we took the effective string coupling $G_s(\mu)$ dependent on the energy scale $\mu$ and following \cite{Seiberg:1999vs} considered the case of abelian gauge field, to simplify the derivation of the effective YM coupling. To take into account the running of the coupling we compare the effective action (\ref{BI action}) with the DBI action of a probe Dp--brane at fixed radial distance $r(\mu)=r(\rho(\mu))$. Given that we have $1+3$ dimensional field theory we consider probe D3--brane extended along the field theory directions. It is easy to show that such a brane can be stabilized at any constant $r$. The corresponding DBI action is
\begin{equation}\label{DBI action}
{\cal L}(F)=\frac{1}{(2\pi)^3\alpha'^2}\,e^{-\Phi_s({r(\mu)})}\, \sqrt{\det(g+B+2\,\pi\,\alpha'\,F)}\ ,
\end{equation}
where the dilaton is given by equation (\ref{dilaton-string}) and we take $g$ to be the induced metric on the worldvolume of the D3--brane. Given that the D3--brane is stable one can take the induced metric in the static gauge, when it is given by the components of the 10D metric (in the string frame) along the field theory direction. The components of $g$ are then given by equation (\ref{G-string}). Following \cite{Seiberg:1999vs} we compare the actions (\ref{BI action}) and (\ref{DBI action}) at $\hat F=0$ and $F=0$. Setting $p=3$ in  (\ref{BI action}) for the effective string coupling we obtain
\begin{equation}
G_s(\mu)= e^{\Phi_s(r(\mu))}\,\left(\frac{\det(g+B)}{\det g} \right)^{1/2}=e^{\Phi_0(\mu)}= \frac{g_s}{Q_f\,\log\frac{\Lambda_{LP}}{\mu}}=\frac{2\pi}{N_f}\, \frac{1}{\log\frac{\Lambda_{LP}}{\mu}}\ ,
\end{equation}
where we have used equations (\ref{Solution-Z}), (\ref{LP}), (\ref{G-string}), (\ref{dilaton-string}) and (\ref{radius/energy}). Note that the effective string coupling constant does not depend on the parameter of non-commutativity. The effective YM coupling can be calculated by obtaining the coefficient in front of the term $\frac{1}{4}\,G^{ii'}\,G^{jj'}\, \hat F_{ij}\ast \hat F_{ij}$ in the small $\hat F$ expansion of (\ref{BI action}). Our final expression for the effective YM coupling is
\begin{equation}
g_{YM}^2\,= \,\frac{4\pi^2}{N_f}\,\frac{1}{\log{\frac{\Lambda_{LP}}{\mu}}}\ ,
\end{equation}
which is independent of $\Theta$ and is the same as in the commutative limit.

It is an easy exercise to calculate the corresponding beta function
\begin{equation}
\beta_{g_{YM}^2}\,=\,\frac{\partial\,g_{YM}^2}{\partial\,(\log\frac{\mu}{\Lambda_{LP}})} =\frac{4\,\pi^2}{N_f}\,\frac{1}{\left(\log\frac{\mu}{\Lambda}\right)^2} >0\ ,
\end{equation}
which is positive at all energy scales  consistent with the presence of a UV Landau pole and the vanishing of the YM coupling in the deep IR.


\section{Discussion}

In this paper we constructed a supergravity background dual to ${\cal N}=1$ Non-Commutative Super Yang Mills theory with massless smeared flavors. Our solution generalizes the supergravity background dual to the ordinary flavored ${\cal N}=1$ SYM obtained in \cite{Bigazzi:2009bk}. Close to the origin our background reduces to the background of \cite{Bigazzi:2009bk}, reflecting that in the IR limit the NCSYM theory reduces to an ordinary SYM. At large radial distance the dilaton diverges signaling the presence of a UV Landau pole in the dual gauge theory. 

Using the approach of \cite{Seiberg:1999vs} we considered a probe D3-brane at fixed radial distance and extended along the field theory directions. Comparing the corresponding DBI action to an effective BI action for the non-commutative SYM theory, we calculated the effective YM coupling $g_{YM}^2$. Our results show that the effective coupling does not depend on the parameter of non-commutativity and coincides with the expression in the ordinary case. We calculated the corresponding beta function showing that it is positive, which is consistent with the existence of UV Landau pole and the fact that in the IR the gauge coupling vanishes.

As mentioned above the near horizon region of our background is the same as in the ordinary case studied in \cite{Bigazzi:2009bk}. In the Einstein frame the near horizon region is AdS$_5\times S^5$, however the dilaton is not constant and $e^{\phi} \to 0$ near the origin. This leads to a curvature singularity in the string frame, which is an artifact of the smearing of the flavor branes and can be removed by giving a small mass to the flavors. Indeed, in \cite{Bigazzi:2009bk} and \cite{Bigazzi:2008zt} it was shown that the finite mass of the flavors is reflected in the dual supergravity background to the presence of a spherical cavity with radius proportional to the mass of the flavors. Inside the cavity the solution is given by the vacuum solution (without flavors). The vacuum solution has a constant dilaton in the IR and no curvature singularities in the string frame. 
Preliminary studies in the massive flavor case \cite{work in progress} show that in analogy to \cite{Bigazzi:2009bk} and \cite{Bigazzi:2008zt}, the geometry develops a spherical cavity of radius proportional to the mass of the flavors, but with a vacuum solution given by the Maldacena-Russo background \cite{Hashimoto:1999ut, Maldacena:1999mh}.

Let us close our discussion by outlining some directions for future studies. One of the obvious directions, which is already in progress, is the construction of the supergravity background dual to NCSYM with massive flavors. As discussed above, in the massive case the dilaton remains constant inside the cavity suggesting that the coupling constant doesn't vanish in the IR, which makes the theory phenomenologically more relevant. It would be interesting to study the behavior of the beta function of the theory with massive flavors. Another possible direction is the study of the meson spectrum of the theory by introducing probe $D7$-branes. It would be interesting to compare with the meson spectrum in the quenched approximation, performed in \cite{Arean:2005ar}.

\section{Acknowledgements}

We would like to thank D. O'Connor, S. Kovacs, C.  N\'u\~nez and especially A.V. Ramallo for their useful comments and suggestions.
The research of G. Itsios has been co-financed by the ESF
and Greek national funds through the Operational Program ``Education and Lifelong Learning'' of the NSRF - Research Funding Program: ``Heracleitus II.
 Investing in knowledge in society through
the European Social Fund''.
The work of V. F. was supported by an INSPIRE IRCSET-Marie Curie International Mobility Fellowship.
D.~Z.~is funded by the FCT fellowship SFRH/BPD/62888/2009.
Centro de F\'{i}sica do Porto is partially funded by FCT through the projects
PTDC/FIS/099293/2008 and CERN/FP/116358/2010.


\appendix 

\section{Analysis of the BPS equations} \label{BPSanalysis}

In this section we will present the analytic derivation of the system of BPS equations that appears in \eqref{BPS-b}-\eqref{BPS-F}. 
The supersymmetric transformations of the dilatino $\lambda$ and the gravitino  $\psi_\mu$ in type IIB supergravity (in the Einstein frame) \cite{Benini:2006hh},
that we will use for our analysis are the following
\begin{eqnarray} \label{dil&grav}
\delta_{\epsilon}  \lambda&=&{1 \over 2}\Gamma^M \Big[  \partial_M\phi-e^{\phi}F_M^{(1)}(i\sigma_2) \Big]
\epsilon\,- \,{1 \over 4}{1 \over 3!} \Gamma^{MNP}
\Big[e^{-{\phi \over 2}} H_{MNP}\sigma_3 \, + \,  e^{{\phi \over 2}}F_{MNP}^{(3)} \sigma_1 \Big]\epsilon   \, ,  
\nonumber \\ [3pt]
\delta_{\epsilon}  \psi_M&=&\nabla_M\epsilon+{1 \over 4}e^{\phi}F_M ^{(1)}(i\sigma_2)\epsilon \, + \, 
{1 \over 96}\big[e^{-{\phi \over 2}} H_{NPQ}\sigma_3 \, - \,  e^{{\phi \over 2}}F_{NPQ}^{(3)} \sigma_1 \big] \big[ \Gamma_M^{\,\, NPQ}-9\delta^N_M\Gamma^{PQ} \big]\epsilon
\nonumber \\
&+&{1 \over 16}{1 \over 5!}F_{NPQRT}^{(5)}\Gamma^{NPQRT}(i\sigma_2)\Gamma_M\epsilon \,\, ,
\end{eqnarray}
where $\Gamma^{M}$are the ten-dimensional Dirac matrices, $\sigma_i \,\, i=1,2,3$ are the Pauli matrices and $\epsilon$ is a complex Weyl spinor 
of fixed ten-dimensional chirality.

\noindent We start the analysis with the dilatino variation in \eqref{dil&grav}. We can derive
a differential equation for the dilaton using the expressions for the differential forms in \eqref{H3F1F3frame} and \eqref{F5frame}, while demanding 
the dilatino variation to vanish
\begin{eqnarray} \label{eq:dil1}
&& \Phi'(\sigma) \, \epsilon \, - \, Q_f \, e^{\Phi(\sigma)} \, b(\sigma) \, S(\sigma)^4 \, \Gamma^{49} (i\sigma_2) \, \epsilon \, - \, 
\frac{1}{2} \, e^{- {1 \over 2}\, \Phi(\sigma)} \, h(\sigma)^{1/2} \, b(\sigma)^{-1} \, H'(\sigma) \, \Gamma^{23} \sigma_3 \, \epsilon 
\nonumber \\ [3.5pt]
&& - \frac{1}{2} \, e^{{1 \over 2}\, \Phi(\sigma)} h(\sigma)^{1/2} \, J'(\sigma) \, \Gamma^{01} \, \sigma_1 \, \epsilon \, - \, 
\frac{1}{2} \, Q_f \, e^{{1 \over 2}\, \Phi(\sigma)} \, h(\sigma)^{1/2} \, S(\sigma)^{4} H(\sigma) \, \Gamma^{2349} \sigma_1 \epsilon =0 \, . 
\end{eqnarray}
Next, we analyze the gravitino variation (\ref{dil&grav}) along the directions $t, x_1, x_2,x_3$ \& $\sigma$.
The components of the spin connection essential for this analysis are 
\begin{equation}
\begin{array}{l}
\omega_{0}^{\,\, 04} \, = \, \omega_{1}^{\,\, 14} \, = \, - \, \frac{1}{4} \, h'(\sigma) \, h(\sigma)^{-5/4} \, b(\sigma)^{-1} \, F(\sigma)^{-1} \, S(\sigma)^{-4} \, ,\\[10pt]
\omega_{2}^{\,\, 24} \, = \, \omega_{3}^{\,\, 34} \, = \, \frac{1}{4} \Big[2 h(\sigma) \, b'(\sigma) \, - \,  b(\sigma) \, h'(\sigma) \Big]  \, h(\sigma)^{-5/4} \, b(\sigma)^{-2} \, F(\sigma)^{-1} \, S(\sigma)^{-4} \, .
\end{array}
\end{equation}
Putting all the ingredients in \eqref{dil&grav}, we arrive to the following set of differential equations
\\
\\
\noindent $\bullet$ Along the directions $M=0, 1$:
\\
\begin{equation}
\label{eq:gravx1}
\begin{aligned}
& h'(\sigma) \, \epsilon \, - \,\frac{1}{2} \, e^{- {1 \over 2}\, \Phi(\sigma)} \,  H'(\sigma) h(\sigma)^{3/2} \, b(\sigma)^{-1} \, \Gamma^{23} \, \sigma_3 \, \epsilon \, - \, 
Q_{c} \, b(\sigma)\,  \Gamma^{0123} (i\sigma_2) \, \epsilon \\[5pt]
& - \, \frac{3}{2} \, e^{{1 \over 2}\, \Phi(\sigma)}\, h(\sigma)^{3/2} \, J'(\sigma) \, \Gamma^{01} \, \sigma_1 \, \epsilon \, + \, 
\frac{1}{2} \, e^{ {1 \over 2}\, \Phi(\sigma)} Q_{f} \, H(\sigma) \, h(\sigma)^{3/2} \, S(\sigma)^4 \, \Gamma^{2349} \, \sigma_1 \, \epsilon = \,0
 \end{aligned}
\end{equation}
\\
\noindent $\bullet$ Along the directions $M=2, 3$: 
\\
\begin{equation}
\label{eq:gravx2}
\begin{aligned}
 & \Big[ h'(\sigma) \, - \,  2 \, h(\sigma) \, \partial_{\sigma} \log b(\sigma) \Big] \, \epsilon \, + \, 
\frac{1}{2} \, e^{ {1 \over 2}\, \Phi(\sigma)}\,  J'(\sigma) \, h(\sigma)^{3/2} \, \Gamma^{01} \, \sigma_{1} \, \epsilon \, - \, 
Q_{c} \, b(\sigma) \, \Gamma^{0123} (i\sigma_2) \, \epsilon\\[5pt]
 & - \,\frac{3}{2} \, e^{{1 \over 2}\Phi(\sigma)} \, Q_{f} \, H(\sigma) \, h(\sigma)^{3/2} \, S(\sigma)^4 \, \Gamma^{2349} \, \sigma_1 \, \epsilon \,+ \, 
\frac{3}{2} \, e^{-\frac{1}{2}\Phi(\sigma)} \, H'(\sigma) \, h(\sigma)^{3/2} \, b(\sigma)^{-1} \, \Gamma^{23} \, \sigma_3 \, \epsilon   = \, 0
 \end{aligned}
\end{equation}

\noindent $\bullet$ Along the direction $M=4$:
\\
\begin{equation}
\label{eq:gravs}
\begin{aligned}
& \partial_{\sigma} \epsilon \, - \, \frac{3}{16} e^{- {1 \over 2}\Phi(\sigma)} \, H'(\sigma) \, h(\sigma)^{1/2} \, b(\sigma)^{-1} \, \Gamma^{23} \, \sigma_1 \, \epsilon \, + \, 
\frac{1}{8} \,Q_{c} \, b(\sigma) \,h(\sigma)^{-1} \, \Gamma^{0123} \, (i\sigma_2) \, \epsilon  \\[5pt]
 &+ \frac{3}{16} e^{{1 \over 2}\Phi(\sigma)} \, J'(\sigma) \, h(\sigma)^{1/2} \, \Gamma^{01} \, \sigma_1 \, \epsilon \, - \, 
{1 \over 16} \, e^{{1 \over 2}\Phi(\sigma)} Q_{f} \, H(\sigma) \, h(\sigma)^{1/2} \, S(\sigma)^4 \, \Gamma^{2349} \, \sigma_1 \, \epsilon \, = \, 0
 \end{aligned}
\end{equation}
Note that we have assumed that the spinor $\epsilon$ is independent of the coordinates $t,x_1,x_2,x_3$. 
The projections we will impose in order to obtain the BPS equations for the different functions of the background are 
\begin{equation} \label{projection-v1}
 i \, \Gamma_{0123}\,\sigma_2 \, \epsilon \, = \,  i \, \Gamma_{49}\,\sigma_2 \, \epsilon \, = \,  e^{- \, \alpha \, \Gamma_{23} \, \sigma_3}\, \epsilon \, .
\end{equation}
Combining \eqref{eq:gravx1} and \eqref{eq:gravx2} we have
\begin{equation}
\label{eq:gravx1x2}
\begin{aligned}
& \partial_{\sigma} \log{b(\sigma)}\epsilon \, - 
\, e^{-\, {1 \over 2} \Phi(\sigma)} \, H'(\sigma) \, h(\sigma)^{1/2} \, b(\sigma)^{-1} \, \Gamma^{23} \,\sigma_{3} \, \epsilon  \\[8pt]
& + \,  e^{{1 \over 2} \,\Phi(\sigma)} \, Q_{f} \, H(\sigma) \, h(\sigma)^{1/2} \, S(\sigma)^4 \, \Gamma^{2349} \, \sigma_1 \, \epsilon
\, - \, e^{{1 \over 2}\Phi(\sigma)} \,J'(\sigma) \, h(\sigma)^{1/2} \, \Gamma^{01} \, \sigma_1 \, \epsilon \, = \, 0 \, .
\end{aligned}
\end{equation}
Imposing \eqref{projection-v1} to the previous equation and collecting terms proportional to $\mathbbm{1}$ and to $\Gamma^{23}$,
we obtain \eqref{BPS-b} and the following equation respectively
\begin{equation} \label{eq:derH1}
H'(\sigma) \, - \, e^{\Phi(\sigma)} \, Q_{f} \, b(\sigma) \, H(\sigma) \, S(\sigma)^4 \, \cos\alpha(\sigma) \, + 
\, e^{\Phi(\sigma)} \, J'(\sigma) \, b(\sigma) \, \cos\alpha(\sigma) \, = \, 0 \, . 
\end{equation}
Combining  \eqref{eq:dil1}, \eqref{eq:gravx1x2}, and \eqref{projection-v1} and collecting terms proportional to $\Gamma^{23}$ and $\mathbbm{1}$,
we obtain \eqref{tana} and \eqref{BPS-Phi} respectively.
Another set of equations can be derived by imposing the projections to \eqref{eq:gravx1}. Terms proportional to $\mathbbm{1}$ will give us \eqref{BPS-h}
while terms proportional to $\Gamma^{23}$ will give us the following equation
\begin{equation} \label{eq:derH2}
\begin{aligned}
& H'(\sigma) \, - \, e^{\Phi(\sigma)} \, Q_{f} \, b(\sigma) \, H(\sigma) \, S(\sigma)^4 \, \cos{\alpha(\sigma)}
\\[5pt]
&+ \, 2\, Q_{c}\,  e^{{1 \over 2} \, \Phi(\sigma)} b(\sigma)^2 \, h(\sigma)^{-3/2} \, \sin{\alpha(\sigma)}\,  + \, 
3 \, e^{\Phi(\sigma)} \, J'(\sigma) \, b(\sigma) \, \cos{\alpha(\sigma)} \, = \, 0 \, . 
 \end{aligned}
\end{equation}
Subtracting  \eqref{eq:derH2} from \eqref{eq:derH1} and using (\ref{tana}) we end up with \eqref{BPS-J} and \eqref{BPS-H}.
Inserting all the ingredients in the gravitino equation along the radial direction and collecting terms
 proportional to the identity matrix, we arrive at
\begin{equation}
\partial_{\sigma} \hat{\epsilon}  \, + \, \frac{1}{8} \, \partial_{\sigma}\log h(\sigma) \, \hat{\epsilon}\, = \, 0 \quad {\rm with} \quad
\epsilon \, = \,  e^{\frac{\alpha}{2} \, \Gamma_{23} \, \sigma_3} \, \hat{\epsilon} \, .  
\end{equation}
We can integrate the above equation to obtain an expression for the spinor $\hat{\epsilon}$
\begin{equation}
\label{eq:spinorhat}
 \hat{\epsilon} = h(\sigma)^{-1/8} \tilde{\epsilon} \, ,
\end{equation}
where $\tilde{\epsilon}$ does not depend on the coordinates $t,x_1,x_2,x_3,\sigma$.
\\
\noindent Next, we consider the gravitino equation along the flat directions $M=5,6,7,8$ \& $9$ and there some extra projections we need to impose 
\begin{equation}  \label{projection-v2}
\Gamma^{58} \, \epsilon \,  = \,  \Gamma^{67} \, \epsilon \,  =  \, - \, \Gamma^{49} \, \epsilon  \, .
\end{equation}

\noindent $\bullet$ Along the direction $M=5$:
\\
\begin{equation}
\label{eq:direction5}
\begin{aligned}
& \partial_{\sigma} \log\left[ h(\sigma) \, S(\sigma)^4 \right] \, \epsilon \, - \,  4\,  b(\sigma) \, F(\sigma)^2 \, S(\sigma)^2 \, \Gamma^{4589}\, \epsilon \, - \, 
Q_{c} \, b(\sigma) \, h(\sigma)^{-1} \, i \sigma_{2} \, \Gamma^{0123} \, \epsilon \\[5pt]
  & - \,  \frac{1}{2} \, e^{\frac{1}{2} \, \Phi(\sigma)} \, J^{\prime}(\sigma)\,  h(\sigma)^{1/2} \, \Gamma^{01} \,  \sigma_{1} \, \epsilon \,  - \,  
\frac{1}{2} \, Q_{f} \, e^{\frac{1}{2} \, \Phi(\sigma)} \, h(\sigma)^{1/2} \, H(\sigma) \, S(\sigma)^{4}  \, \Gamma^{2349} \, \sigma_{1} \, \epsilon \\[5pt]
  & +\, \frac{1}{2} \, e^{-\frac{1}{2} \, \Phi(\sigma) } \, H^{\prime}(\sigma)\,  h(\sigma)^{1/2} \, b(\sigma)^{-1} \, \Gamma^{23} \, \sigma_{3} \, \epsilon  \, = \, 0 \, .
 \end{aligned}
\end{equation}
To obtain the above result we have considered that the spinor does not depend on the coordinate $\chi$, while we have also used the
following components of the spin connection
\begin{equation}
 \begin{aligned}
  & \omega_{5}^{ \, \, 45} \, =\,  -\, \frac{1}{4} \, \partial_{\sigma} \log\left[ h(\sigma) \, S(\sigma)^4 \right] \, 
b(\sigma)^{-1} \, F(\sigma)^{-1}\,  h(\sigma)^{-1/4} \, S(\sigma)^{-4} \, , 
\\[5pt]
  & \omega_{5}^{ \, \, 89} \, = \,  - \,  F(\sigma) \,  h(\sigma)^{-1/4} \, S(\sigma)^{-2} \, .
 \end{aligned}
\end{equation}
Combining  \eqref{projection-v1}, \eqref{projection-v2}, \eqref{eq:gravx1} \& \eqref{eq:direction5} we obtain  \eqref{BPS-S}.
\\
\\
\noindent $\bullet$ Along the direction $M=6$:
\\
\begin{equation}
\label{eq:direction6}
 \begin{aligned}
& \partial_{\sigma} \log\left[ h(\sigma) \, S(\sigma)^4 \right] \, \epsilon \, - \,  4\,  b(\sigma) \, F(\sigma)^2 \, S(\sigma)^2 \, \Gamma^{4679}\, \epsilon \, - \, 
Q_{c} \, b(\sigma) \, h(\sigma)^{-1} \, i \sigma_{2} \, \Gamma^{0123} \, \epsilon 
\\[5pt]
  & - \,  \frac{1}{2} \, e^{\frac{1}{2} \, \Phi(\sigma)} \, J^{\prime}(\sigma)\,  h(\sigma)^{1/2} \, \Gamma^{01} \,  \sigma_{1} \, \epsilon \,  - \,  
\frac{1}{2} \, Q_{f} \, e^{\frac{1}{2} \, \Phi(\sigma)} \, h(\sigma)^{1/2} \, H(\sigma) \, S(\sigma)^{4}  \, \Gamma^{2349} \, \sigma_{1} \, \epsilon 
\\[5pt]
  & +\, \frac{1}{2} \, e^{-\frac{1}{2} \, \Phi(\sigma) } \, H^{\prime}(\sigma)\,  h(\sigma)^{1/2} \, b(\sigma)^{-1} \, \Gamma^{23} \, \sigma_{3} \, \epsilon  \, 
- \, 16 \,  b(\sigma) \, F(\sigma) \, S(\sigma)^3 \, \cos^{-1}\frac{\chi}{2}\, \Gamma^{64}\, \partial_{\theta} \, \epsilon  
 \\[5pt] &
- \, 4 \, b(\sigma) \, F(\sigma) \, S(\sigma)^3 \, \tan\frac{\chi}{2}\, \Gamma^{64}\, \left[ \Gamma^{56} \, - \, \Gamma^{78}\right] \, \epsilon   \, = \, 0 \, .
 \end{aligned}
\end{equation}
For the above result,  the following components of the spin connection are needed
\begin{equation}
 \begin{aligned}
  & \omega_{6}^{ \, \, 46} \,  = \, - \, \frac{1}{4} \, \partial_{\sigma} \, \log\left[ h(\sigma) \, S(\sigma)^4\right] \, 
b(\sigma)^{-1} \, F(\sigma)^{-1} \, h(\sigma)^{-1/4} \, S(\sigma)^{-4} \, , \\
  & \omega_{6}^{ \, \, 56} \, = \,  -  \, \omega_{6}^{ \, \, 78} \, = \,  h(\sigma)^{-1/4} \, S(\sigma)^{-1} \, \tan \frac{\chi}{2} \, ,  \quad \& \quad
\omega_{6}^{ \, \, 79} \, = \,  F(\sigma)  \, h(\sigma)^{-1/4}  \, S(\sigma)^{-2} \, .
 \end{aligned}
\end{equation}
Combining \eqref{projection-v2}, \eqref{eq:direction5} \& \eqref{eq:direction6} we conclude that the spinor does not depend on the coordinate 
$\theta$, i.e $\partial_{\theta} \, \epsilon \, =  \, 0$.
\\
\\
\noindent $\bullet$ Along the direction $M=7$:
\\
\\
The relevant spin connection components are the following
\begin{equation}
 \begin{aligned}
  & \omega_{7}^{ \, \, 47} \, = \, - \, \frac{1}{4} \partial_{\sigma} \log\left[ h(\sigma) \, S(\sigma)^4\right] \, 
  b(\sigma)^{-1} \, F(\sigma)^{-1} \, h(\sigma)^{-1/4} \, S(\sigma)^{-4} \, ,\\[5pt]
  & \omega_{7}^{ \, \, 57} \, = \, \omega_{7}^{ \, \, 68} \, = \,  h(\sigma)^{-1/4} \, S(\sigma)^{-1} \, \tan \frac{\chi}{2} \, , \\[5pt]
  & \omega_{7}^{ \, \, 67} \, = \, - \,  2 h(\sigma)^{-1/4} \, S(\sigma)^{-1} \, \cot\theta \, \cos^{-1}\frac{\chi}{2} \, , \quad \& \quad 
  \omega_{7}^{ \, \, 69} \, = \,  F(\sigma) \, h(\sigma)^{-1/4} \, S(\sigma)^{-2} \, .
 \end{aligned}
\end{equation}
Combining \eqref{eq:direction5} \& \eqref{eq:direction6} and assuming that the 
spinor is independent of the coordinate $\phi$, i.e $\partial_{\phi}\, \epsilon \, = \,  0$, we arrive to the following differential equation
\begin{equation}
\label{eq:spinortilde}
 \partial_{\psi} \tilde{\epsilon} \, +  \, \frac{1}{2} \, i \, \sigma_2 \, \tilde{\epsilon} \, = \,  0 \quad \Rightarrow \quad 
\tilde{\epsilon} \, = \,  e^{-\frac{1}{2} \, i\, \sigma_2 \, \psi} \, \epsilon_0 \, . 
\end{equation}

\noindent $\bullet$ Along the direction $M=8$:
\\
\\
The relevant spin connection components are the following
\begin{equation}
 \begin{aligned}
  & \omega_{8}^{ \, \, 48} \, = \,  - \, \frac{1}{4} \, \partial_{\sigma} \, \log\left[ h(\sigma) \, S(\sigma)^4 \right] \,  b(\sigma)^{-1} \, F(\sigma)^{-1} \, h(\sigma)^{-1/4} 
  \, S(\sigma)^{-4} \, , \\[5pt]
  & \omega_{8}^{ \, \, 58} \, = \,  - \, 2 \, S(\sigma)^{-1} \, h(\sigma)^{-1/4} \, \cot\chi \, ,\\[5pt]
  & \omega_{8}^{ \, \, 59} \, = \,  F(\sigma) \, h(\sigma)^{-1/4} \, S(\sigma)^{-2} \, , \quad \& \quad 
  \omega_{8}^{ \, \, 67} \, = \, h(\sigma)^{-1/4} \, S(\sigma)^{-1} \, \tan \frac{\chi}{2} \, .
 \end{aligned}
\end{equation}
Combining \eqref{eq:direction5}, \eqref{eq:direction6} \&  \eqref{eq:spinortilde}, 
we arrive to the following differential equation
\begin{equation}
\label{eq:spinor0}
 \partial_{\tau} \epsilon_0 \, + \,  \frac{3}{2} \, i \, \sigma_2 \, \epsilon_0  \, = \, 0 \quad \Rightarrow \quad
 \epsilon_0 \, = \,  e^{- \frac{3}{2} \, i \, \sigma_2 \, \tau} \eta\, , 
\end{equation}
where $\eta$ is a constant spinor.
\\
\\
\noindent $\bullet$ Along the direction $M=9$:
\\
\begin{equation}
\label{eq:direction9}
 \begin{aligned}
& \partial_{\sigma} \log\left[ h(\sigma) \, F(\sigma)^4 \right] \, \epsilon \, 
+\, \frac{1}{2} \, e^{-\frac{1}{2} \, \Phi(\sigma) } \, H^{\prime}(\sigma)\,  h(\sigma)^{1/2} \, b(\sigma)^{-1} \, \Gamma^{23} \, \sigma_{3} \, \epsilon
\\[5pt] 
& - \, 
Q_{c} \, b(\sigma) \, h(\sigma)^{-1} \, i \sigma_{2} \, \Gamma^{0123} \, \epsilon \, +  \, 
2 \, Q_f \, e^{\Phi(\sigma)} \, b(\sigma) \, S(\sigma)^4 \, \Gamma^{49} (i\sigma_2) \, \epsilon  
\\[5pt]
  & - \,  \frac{1}{2} \, e^{\frac{1}{2} \, \Phi(\sigma)} \, J^{\prime}(\sigma)\,  h(\sigma)^{1/2} \, \Gamma^{01} \,  \sigma_{1} \, \epsilon \, 
+\,  4\,  b(\sigma) \, F(\sigma)^2 \, S(\sigma)^2 \, \Gamma^{49}\, \left[ \Gamma^{58} \, + \, \Gamma^{67}\right] \epsilon
 \\[5pt] &
+ \, \frac{3}{2} \, Q_{f} \, e^{\frac{1}{2} \, \Phi(\sigma)} \, h(\sigma)^{1/2} \, H(\sigma) \, S(\sigma)^{4}  \, \Gamma^{2349} \, \sigma_{1} \, \epsilon 
 \,+ \, 8 \, b(\sigma) \, S(\sigma)^4 \, \Gamma^{49} \, \partial_{\tau} \epsilon  \, = \, 0 \, .
 \end{aligned}
\end{equation}
For the above result,  the following components of the spin connection are needed
\begin{equation}
 \begin{aligned}
   & \omega_{9}^{ \, \, 49} \,  = \, - \, \frac{1}{4} \, \partial_{\sigma} \, \log\left[ h(\sigma) \, F(\sigma)^4\right] \, b(\sigma)^{-1} \, F(\sigma)^{-1} \, h(\sigma)^{-1/4} \, 
S(\sigma)^{-4} \, ,\\[5pt]
   & \omega_{9}^{ \, \, 58} \, = \,  \omega_{9}^{ \, \, 67} \, = \,  F(\sigma) \, h(\sigma)^{-1/4} \, S(\sigma)^{-2} \, .
 \end{aligned}
\end{equation}
Combining \eqref{projection-v1}, \eqref{projection-v2}, \eqref{eq:spinor0}, \eqref{eq:gravx1} \& \eqref{eq:direction9} we obtain \eqref{BPS-F}.


\section{Equations of motion} \label{EOMs}

The equations of motion for every function that appears in the background is
\begin{eqnarray}
\partial_\sigma^2 \log b(\sigma) &=& 
-\, \frac{4 \, Q_f\,H^2\, h \, S^6 \, F^2}{\beta_1}
\,-\,Q_f^2\, e^{\Phi}\,H^2\, h \,S^8\,\beta_3 \, , 
\label{diff-b} \\ [4pt]
\partial_\sigma^2 \log h(\sigma)&=& 
- \,Q_c^2 \,\frac{b^2}{h^2}\,-\, \frac{2 \,Q_f\,H^2\, h \, S^6 \, F^2}{\beta_1}\,-\,
\frac{1}{2}\,Q_f^2 \,e^{\Phi}\,H^2\, h \,S^8\,\beta_3
\nonumber \\ [4pt]
&+& 
\left(1 \, - \, \beta_2\right)\,\frac{e^{-\Phi}\,h\,H'^2}{b^2} \, , 
\label{diff-h} \\ [4pt]
\partial_{\sigma} \left[\frac{e^{-\Phi(\sigma)}\,h(\sigma)\,H'(\sigma)}{b(\sigma)^2}\right]&=& 
e^{\Phi}\,Q_f^2\,H\,h\,S^8\,-\,Q_c\,J' \,+\, \frac{4 \, Q_f\,H\, h \, S^6 \, F^2}{\beta_1} \, ,
\label{diff-H} \\ [4pt]
\partial_\sigma^2\Phi(\sigma)&=& 
\frac{1}{2}\,\left(1 \,+ \, \beta_1^2 \right)
\left[Q_f^2\,e^{2\Phi}\,b^2\,S^8 \,+\, \frac{4 \, Q_f\,b^2\, e^\Phi \, S^6 \, F^2}{\beta_1}\right]\,
\nonumber \\ [4pt]
&-& \frac{1}{2}\,\frac{e^{-\Phi}\,h\,H'^2\,\beta_2}{b^2} \, , 
\label{diff-Phi} \\ [4pt]
\partial_\sigma^2 \log S(\sigma)&=& 
- \, 2\, b^2 \,  F^4 \, S^4  \,+ \,  6\, b^2 \, F^2 \, S^6 
\nonumber \\ [4pt]
&- & 
\frac{Q_f \,e^\Phi \, b^2 \, F^2\,S^6}{\beta_1} \, + \, 
\frac{1}{4}\,Q_f^2 \, e^{\Phi}\,H^2 \,h \,S^8\, \beta_3 \, ,  
\label{diff-S} \\ [4pt] 
\partial_\sigma^2 \log F(\sigma)&=& 
4\,b^2 \, F^4 \, S^4 \, - \, \frac{1}{4}\,\left(1 \, + \, \beta_1^2 \right)\,Q_f^2\,e^{2\Phi} \, b^2 \, S^8
\nonumber \\ [4pt]
&+&
\frac{Q_f\,H^2\, h \, S^6 \, F^2}{\beta_1} \,+\,\frac{1}{4}\,\frac{e^{-\Phi}\,h \,H'^2\,\beta_2}{b^2}  \, , 
\label{diff-F}
\end{eqnarray}
where we have defined the following (auxiliary) dimensionless expressions
\begin{equation}
\beta_1 \equiv \sqrt{1+\frac{e^{-\Phi}\,H^2\,h}{b^2}} \, , \quad
\beta_2 \equiv 1 + \frac{e^{2\Phi}\,J'^{2}\,b^2}{H'^2} \quad \& \quad
\beta_3 \equiv 1 + \frac{e^{-2\Phi}\,H'^{2}\,\beta_2}{Q_f^2\,H^2\,b^2\,S^8} \, . 
\end{equation}


\end{document}